\documentclass[a4paper,11pt]{article}
\usepackage{pos}
\usepackage{ulem}

\title{Transition Radiation Detector Upgrade for the GlueX--III Charmonium Program}

\author*[a]{Lauren Kasper}
\author[b]{Alexander Austregesilo}
\author[b]{Cody Dickover}
\author[c]{Sean Dobbs}
\author[b]{Sergey Furletov}
\author[b]{Yulia Furletova}
\author[b]{Ilya Larin}
\author[b]{Lubomir Pentchev}
\author[c]{Saheli Rakshit}
\author[c]{Nizar Septian}
\author[b]{Simon Taylor}
\author[a]{Julia Velkovska}

\affiliation[a]{Vanderbilt University, Department of Physics and Astronomy,\\ 
Nashville, TN, 37235, US}
\affiliation[b]{Thomas Jefferson National Accelerator Facility,\\
Newport News, VA, 23606, US}
\affiliation[c]{Florida State University, Department of Physics,\\
Tallahassee, FL, 32306, US}

\emailAdd{lauren.n.kasper@vanderbilt.edu}
\emailAdd{aaustreg@jlab.org}
\emailAdd{dickover@jlab.org}
\emailAdd{sdobbs@fsu.edu}
\emailAdd{furletov@jlab.org}
\emailAdd{yulia@jlab.org}
\emailAdd{ilarin@jlab.org}
\emailAdd{pentchev@jlab.org}
\emailAdd{srakshit@fsu.edu}
\emailAdd{nseptian@fsu.edu}
\emailAdd{staylor@jlab.org}
\emailAdd{julia.velkovska@vanderbilt.edu}

\abstract{The GlueX--III experiment at Jefferson Lab, planned to run in 2027--2029, will enhance the laboratory’s capability to explore charmonium production near threshold and related aspects of QCD in the non--perturbative regime. A central upgrade for GlueX--III is a large-scale triple--GEM Transition Radiation Detector (TRD) optimized for electron identification and pion suppression in the hadron--rich environment of fixed--target photoproduction. This document describes the motivation for and efforts related to such an upgrade, including the design, construction, and performance of a $720\times528$\,mm$^{2}$ triple--GEM--TRD prototype that was tested in the existing \mbox{GlueX--II} experimental acceptance during the 2025 run. Results validate the prototype's integration, timing performance, and electron identification capability. Its projected impact in the GlueX--III setting is discussed, with background reduction in a $J/\psi$ candidate sample demonstrated. The development and use of a modern TRD technology in a high--background environment at Jefferson Lab may inform detector design choices as a future upgrade path for the EIC.}

\FullConference{The 33rd International Workshop on Deep Inelastic Scattering and Related Subjects (DIS2026)\\
4 - 8 May 2026\\
Bologna, Italy\\}

\begin{document}
\maketitle

\section{Introduction and Motivation}

\begin{figure}[!h]
\begin{center}
\includegraphics[width=0.74\textwidth,trim={15pt 10pt 1pt 10pt},clip]{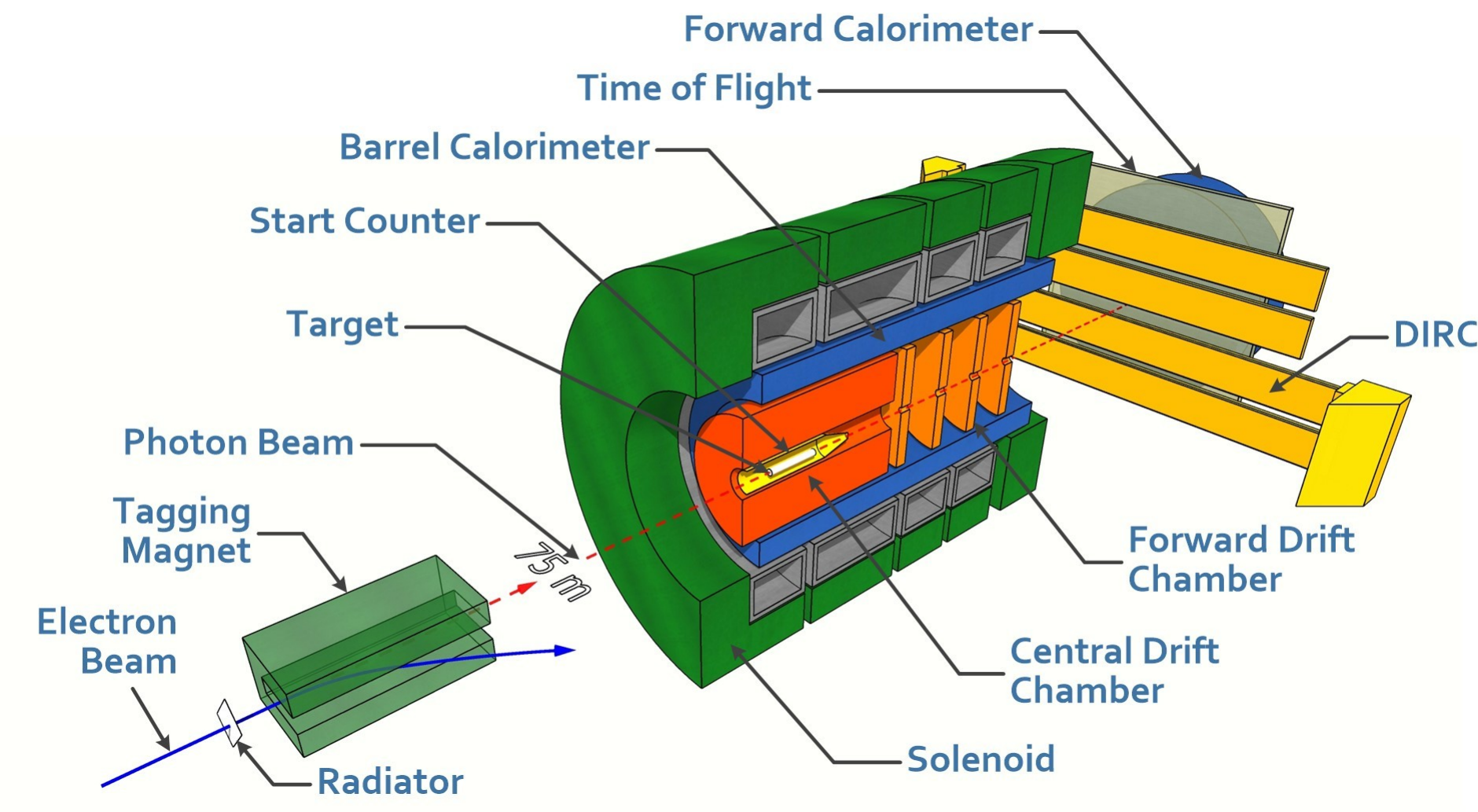}
\caption{Schematic of the GlueX detector layout used as of 2019, highlighting the detector subsystems.}
\label{fig:gluex_schematic}
\end{center}
\end{figure}

The GlueX experiment \cite{gluex} at the Thomas Jefferson National Accelerator Facility (Jefferson Lab) was designed as a spectroscopy effort to map out light-quark mesons up to roughly 3\,GeV/c$^2$. Fig.\ref{fig:gluex_schematic} details the spectrometer's current layout. A superconducting 2\,T solenoid magnet surrounds the target, typically liquid hydrogen, along with central and forward drift chambers and a barrel
calorimeter (BCAL). Time-of-flight (TOF) counters grant particle identification (PID) capability in the forward region, supplemented by a DIRC detector just upstream. Completing the layout is a forward Pb-glass calorimeter (FCAL) with a central PbWO$_4$ electromagnetic calorimeter (ECAL). The spectrometer makes use of a tagged photon beam, produced via coherent bremsstrahlung from 12\,GeV electrons traversing a thin diamond radiator. Special features of GlueX include its nearly 4$\pi$ acceptance and its access to a linear polarization of up to 40\% of the photon beam, achieved in the coherent peak at 9\,GeV.

\begin{figure}[h]
\begin{center}
\includegraphics[width=0.68\textwidth,trim={10pt 0pt 10pt 10pt},clip]{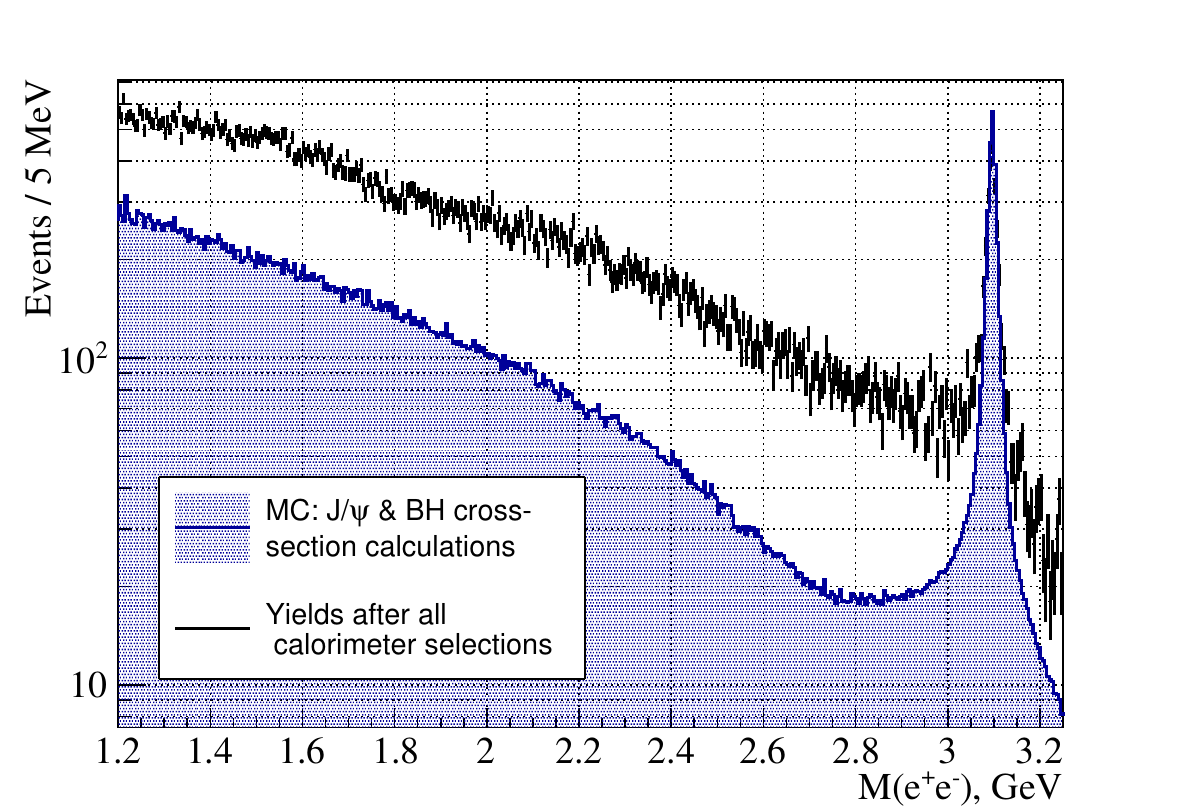}
\caption{Invariant mass spectrum of $e^+ e^-$ from GlueX data (black) compared to Monte Carlo simulations (blue filled) that use absolute BH calculations (as well as the $J/\psi$ photoproduction normalized to the data).}
\label{fig:dataVsMC}
\end{center}
\end{figure} 

The experiment has undergone two main phases: an initial GlueX--I phase at startup and the present GlueX--II phase which includes the DIRC detector upgrade. The currently available datasets from these efforts have produced a wealth of results, the most highly cited of which are on $J/\psi$ photoproduction near-threshold \cite{PRL_Jpsi} \cite{PRC_Jpsi}. Production of charmonium states near-threshold is of interest to numerous physics subfields because it acts as a gluonic probe of Quantum Chromodynamics (QCD) in a uniquely sensitive kinematic regime. Production at threshold is expected to be sensitive to the gluon content of the proton through means like gravitational form factors, or generalized parton distributions \cite{Jpsi_GFFs}; it is also relevant to LHCb pentaquark searches in s-channel \mbox{production \cite{PRL_Jpsi}}.

To further capitalize on the accomplishments of the GlueX spectrometer, Phase-III of the GlueX experiment is planned for 2027-2029. It is intended to run with twice the photon beam intensity and a physics program centered on near-threshold production of charmonium. Phase-III is expected to provide a significant increase in $J/\psi$ statistics as well as access to higher mass $c\overline{c}$ states, such as $\psi(2S)$ and $\chi_{cJ}$. More precise cross section measurements for various $c\overline{c}$ states will serve to motivate the development of a theoretical description of their photoproduction. Current measurements are limited both by statistical precision and residual hadronic contamination of the dielectron sample, restricting the ability to discriminate between proposed charmonium photoproduction mechanisms.

In the case of $J/\psi$ photoproduction, the measured reaction $\gamma p \rightarrow e^+ e^- p$ includes the electromagnetic Bethe-Heitler (BH) process. The BH process provides a calculable continuum that can be used for normalization of the $J/\psi$ cross section measurement. Both processes require sufficient electron identification to combat hadronic background that may mimic the final state $e^+ e^-$ pairs. In the current GlueX measurements, adequate electron identification is provided through selections on the electromagnetic calorimeters. However, existing calorimeter-based electron identification leaves a substantial amount of residual pion contamination beneath the BH continuum as visualized in Fig.~\ref{fig:dataVsMC}. The GEM--TRD upgrade is intended to provide an independent electron-identification observable and suppress this background which dominates the systematic uncertainty of the current $J/\psi$ cross section measurement.

\section{Detector Upgrade Plan}

A transition radiation detector (TRD) is well-suited for the task of electron-hadron ($e/\pi$) separation. The working principle of a TRD is through detection of X-rays, typically 3-50\,keV, emitted by highly relativistic charged particles traversing boundaries between layers of radiator material. Photons are detected via absorption in a heavy-gas mixture, with Xe being the most efficient. TR is emitted by highly relativistic particles ($\gamma=E/m > 1000$), and this $\gamma$-dependence leads to a uniquely broad momentum range where electrons produce it and hadrons do not.

TRDs are a method of low-mass, independent electron identification that allows for systematic studies of uncertainties for other electron ID methods like calorimetry. However, wire-based amplification structures traditionally used in TRDs have suffered from efficiency deterioration due to space-charge effects. An R\&D effort has been underway at Jefferson Lab in recent years to integrate modern high-rate capable MicroPattern Gaseous Detector (MPGD) amplification structures with a TRD. Jefferson Lab scientists have developed and tested GEM-based TRD prototypes and demonstrated proof-of-principle for such a detector, reporting a pion suppression factor of $\sim$$8\pm2$ with small-prototypes \cite{GEM_TRD}. Arising from these efforts is the introduction of a TRD based on GEM amplification, or GEM--TRD, as an additional pion-suppression detector in the GlueX experiment.

\begin{figure}[!h]
\begin{center}
\includegraphics[width=0.68\textwidth,trim={10pt 10pt 2pt 3pt},clip]{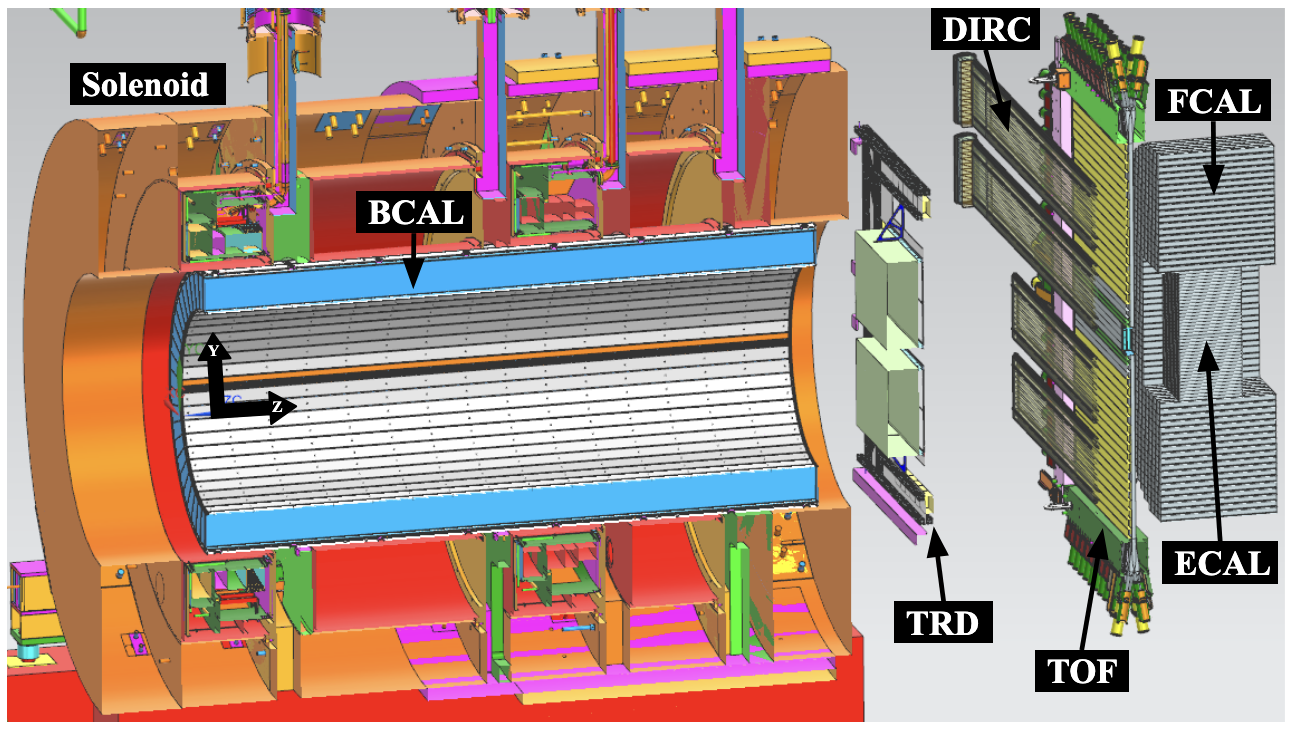}
\includegraphics[width=0.44\textwidth,trim={6pt 3pt 0pt 6pt},clip]{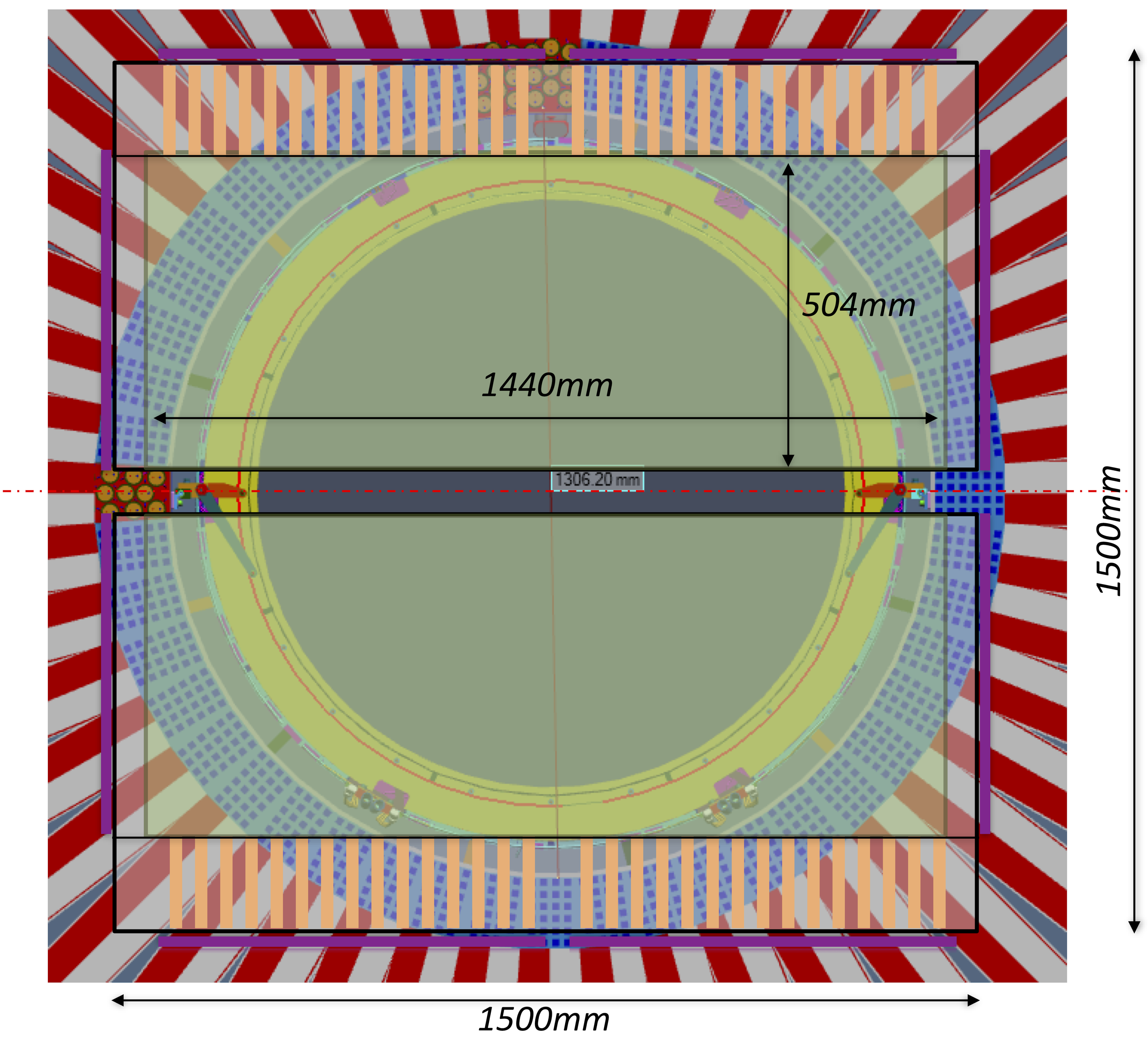}
\includegraphics[width=0.32\textwidth,trim={30pt 30pt 75pt 150pt},clip]{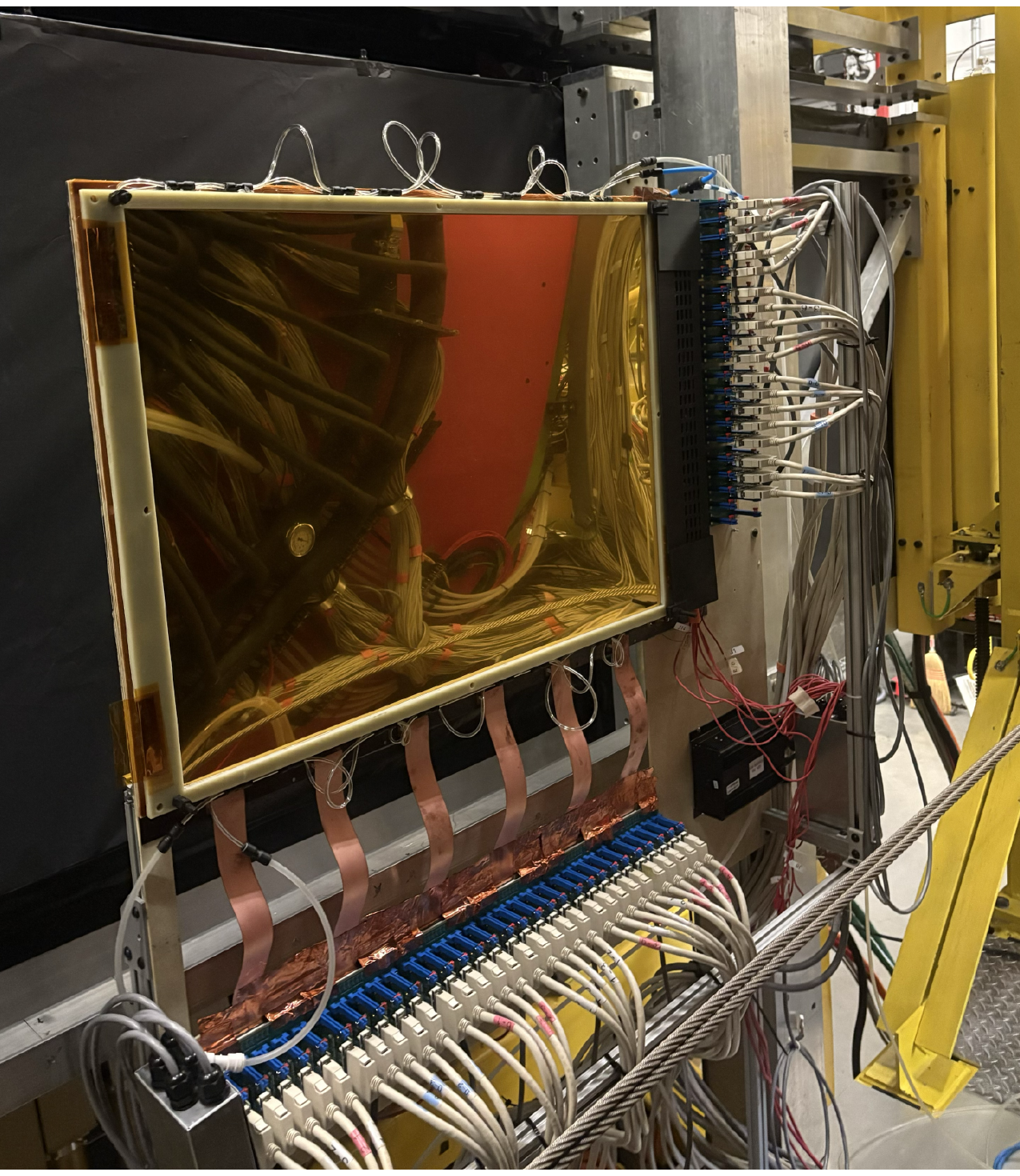}
\caption{(\textit{Top}) Cutaway model of the GlueX--III spectrometer showing the final position in Z of the planned GEM--TRD upgrade; note that GlueX drift chambers are not shown. (\textit{Bottom left}) Upstream view of the acceptance overlap with the GlueX barrel calorimeter for the planned final GEM--TRD. (\textit{Bottom right}) Upstream view of the acceptance overlap with the GlueX barrel calorimeter for the large-scale GEM--TRD prototype as it was installed for the Spring 2025 tests.}
\label{fig:large_detector}
\end{center}
\end{figure} 

The proposed detector is currently being manufactured and is intended to be installed in 2028 for the GlueX--III run. The detector will be installed in the forward region of the GlueX experiment, between the downstream face of the solenoid and the front of the DIRC; its position is shown in Fig.~\ref{fig:large_detector}. The GEM--TRD will consist of two separate chambers, each providing $1440\times504$\,mm$^2$ sensitive area, and a 20\,cm thick radiator layer of fleece-type material. The chambers lie downstream of the radiator and consist of a 2.5\,cm drift volume, three GEM amplification stages separated by 2\,mm transfer gaps, and a readout board. The readout layer consists of X- and Y-strips with capacitive charge sharing, preceded by a Diamond-like carbon layer. The horizontal strips are separated in the middle and are read out from the left and right sides of the chambers. The strip pitch is 1\,mm, resulting in 2448 electronic channels per chamber for 4896 total. The frames of the chambers holding the front-end electronics will lie outside of the experimental acceptance, as displayed in Fig.~\ref{fig:large_detector}.

The detector will utilize a gas collection, purification, and recirculation system to enhance the practicality of Xe gas in a large-scale application. Readout electronics will consist of OpenVPX \cite{VPX} electronics developed for the PANDA experiment and modified for the GlueX GEM--TRD project.

\section{Large--Scale Prototype Tests}

As part of the efforts to realize this detector upgrade, a large-scale GEM--TRD prototype has been recently built with the intent to validate the scalability, integration, and operation of the GEM--TRD concept in the GlueX environment. This large-scale prototype covers a quarter of the final detector (Fig.~\ref{fig:large_detector}) and is constructed with approximately $52\times72$\,cm$^2$ active area. It mirrors the design of the final GlueX--III detector: 3 GEM amplification, 25\,mm drift region, 2\,mm transfer/induction gaps, 2D capacitive sharing strip readout, and 20\,cm of fleece radiator positioned in front of the entrance window. The singular design difference between the two is the entrance window. For the prototype, 25\,\textmu m Kapton is used with 0.1\,\textmu m aluminum adhered as the cathode. The final design features 50\,\textmu m Kapton with 1.5\,\textmu m aluminum adhered.

In Spring of 2025, the large-scale GEM--TRD prototype was installed in the GlueX experimental acceptance downstream of the solenoid. 1080 readout channels were connected to JLab-developed flashADC-125 electronics~\cite{fADC} and integrated with the GlueX data acquisition system (DAQ). Data was collected for about three weeks prior to the production run of GlueX--II. The main purpose of these tests was the study of operational stability and feasibility in terms of the detector's effect on the data rate/volume. Integration with the GlueX DAQ allowed inclusion of the prototype in the GlueX reconstruction software. This enabled the development of three-dimensional charge-cluster reconstruction and track-matching within the detector drift volume.

\begin{figure}[h!]
\begin{center}
\includegraphics[width=0.7\textwidth,trim={750pt 630pt 1090pt 950pt},clip]{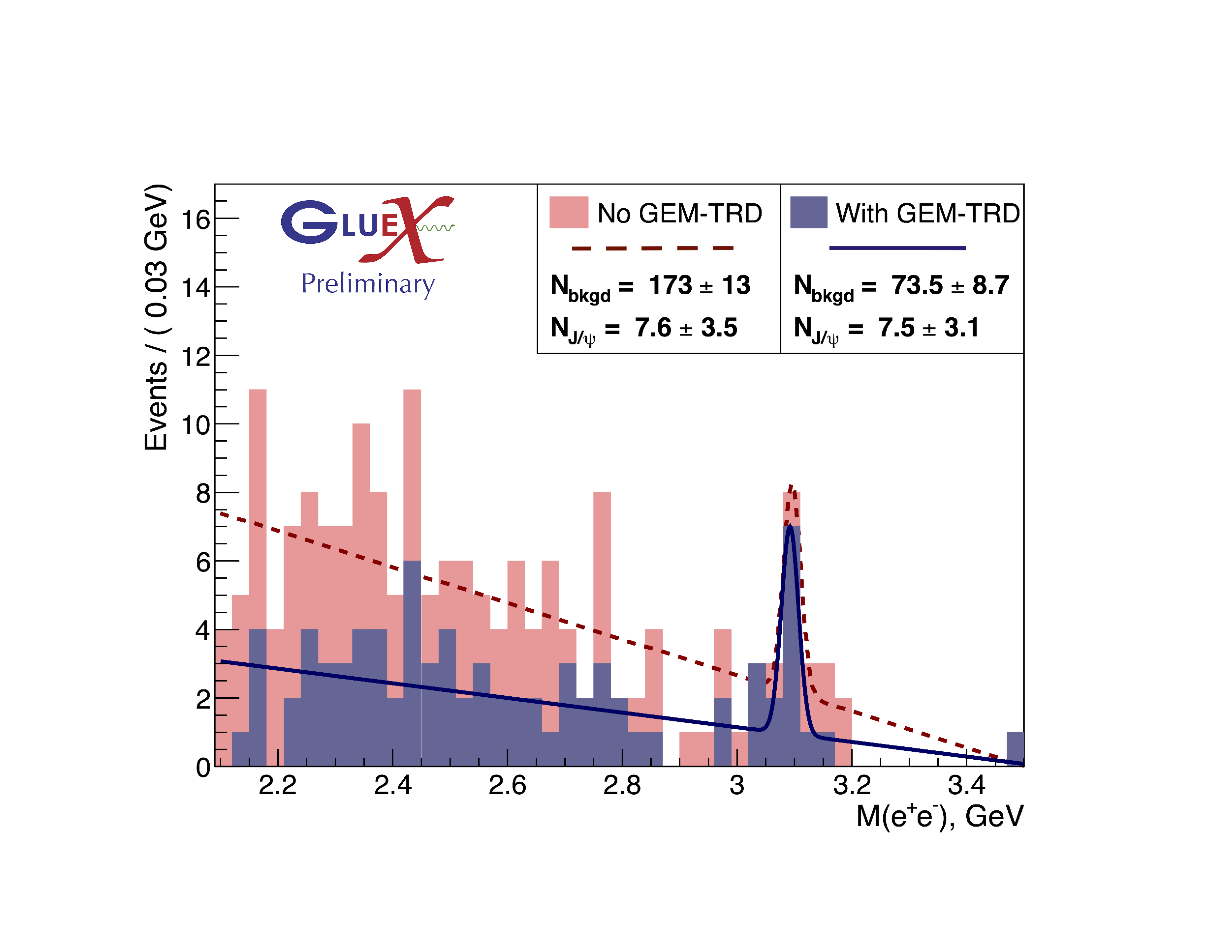}
\caption{Fit of the $e^+e^-$ invariant mass distributions in the region of $2.0 < \text{M}_{e^{+}e^{-}} < 3.6$\,GeV, that includes $J/\psi \rightarrow e^{+}e^{-}$, with and without information from the large-scale GEM--TRD prototype included in the reconstruction process. The solid fill areas represent the data and the lines represent the fit results.}
\label{fig:GEMTRDPeak}
\end{center}
\end{figure}

\begin{figure}[h!]
\begin{center}
\includegraphics[width=0.875\textwidth,trim={20pt 41pt 10pt 25pt},clip]{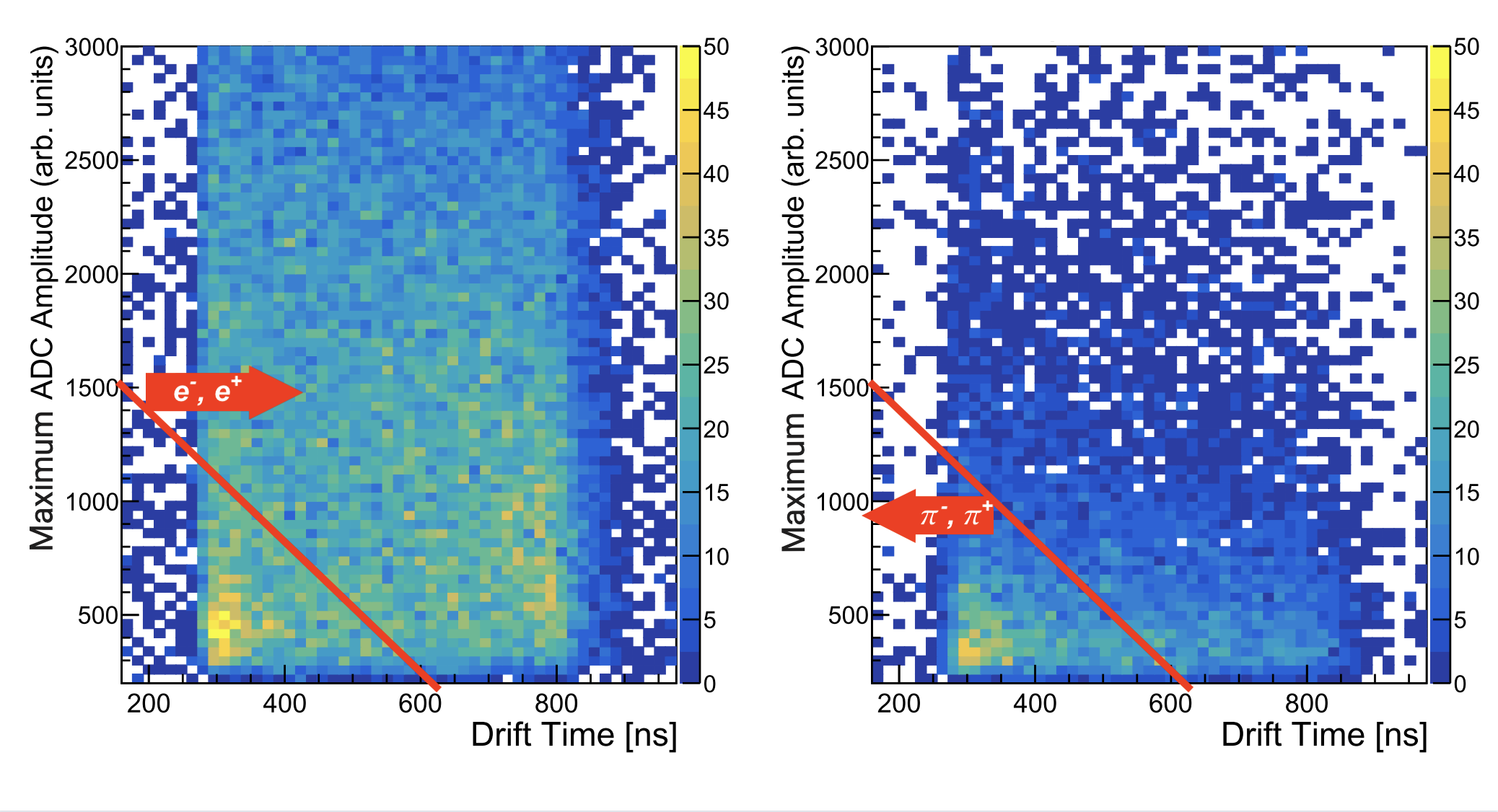}
\caption{Two dimensional distributions of the maximum charge deposit in drift time within the GEM--TRD's drift region for (\textit{left}) $e^{-,+}$ and (\textit{right}) charged hadron samples. The cut used to identify electron signatures corresponds to the distribution: $\text{Amplitude}_{\text{Max}} > 1500 - (2 * \text{DriftTime})$.}
\label{fig:2DAmplitudes}
\end{center}
\end{figure}

These efforts also presented a rare opportunity to evaluate the detector's performance in the actual experimental environment where it will eventually be operated. The prototype was tested with Kr:CO$_2$ (90:10) gas mixture for about twelve days, as a gas collection/recirculation system was not fully developed at the time. Kr is easier to source than Xe, and provides desirable heavy-gas effects necessary for TR applications -- though significantly less efficiently than those of Xe \cite{heavy_gas}.
From this time period, events were skimmed for dielectron final states. By applying $E/p$ cuts to these events ($E$ is the deposited calorimeter energy and $p$ is the reconstructed particle track momentum) $e^{-/+}$ track candidates are identified and a sample is made where at least one of the candidates is projected to have passed through the acceptance of the prototype. The mass distribution of $J/\psi$ candidates is then fitted with a likelihood method after accidental subtraction, shown in Fig.~\ref{fig:GEMTRDPeak}. Events in the `No GEM--TRD' distribution are reconstructed without any information from the GEM--TRD included. Events in the `With GEM--TRD' distribution are reconstructed when the maximum charge deposit in the TRD is selected to be in later time and higher amplitude, as depicted in Fig.~\ref{fig:2DAmplitudes}. Because TR photons are preferentially absorbed near the detector entrance window, their ionization signal arrives later in drift time. Thus, electron events exhibit a higher-amplitude late-time signature; a simple 2D selection can then provide some discrimination power.

The distributions establish that additional PID selection from the GEM--TRD prototype suppresses the background count while preserving the number of $J/\psi $ candidate events. Since operation of the large-scale prototype with Xe was not practical during these tests due to gas recirculation not being possible, in-situ performance with Kr:CO\(_2\) and a singular 2D selection should not be interpreted as the final detector's achievable pion rejection capability. 
Previous small-prototype studies demonstrated approximately a factor 10 pion suppression at 90\% electron efficiency. The final full-size detector for GlueX--III, operated with Xenon as the main gas mixture component, is anticipated to have comparable performance.

\section{Conclusion \& Acknowledgments}

A triple--GEM--TRD enables $e/\pi$ discrimination in high-background environments. Precise $e/\pi$ separation ensures cleaner physics signals and is essential for reducing combinatorial backgrounds in charmonium studies at GlueX--III. The opportunity to test a large-scale prototype of such a detector in the real experimental environment for which one is planned yielded multiple benefits. These tests provided critical insight into the detector operation, directly contributing to the design and commissioning of the final detector for GlueX--III. This novel detector effort also stands to have a broader impact on technology choices for future experiments and facilities, such as the Electron-Ion Collider. There, precise electron identification amidst vast hadronic backgrounds and high rates will be essential for a broad expanse of physics measurements.

This material is based upon work supported by the U.S. Department of Energy, Office of Science, Office of Nuclear Physics under Contract No. 89243126CSC000213 as well as the Office of Workforce Development for Teachers and Scientists, SCGSR program administered by ORISE for the DOE under contract number DE‐SC0014664. This work was also supported in part by Department of Energy Award DE-FG05-92ER40712.

\end{document}